\documentclass[prd,aps,amssymb,nofootinbib,showpacs,preprint]{revtex4}
\usepackage{amsmath,amsfonts}
\usepackage{textcomp,color}
\usepackage{graphics,graphicx}
\usepackage{hyperref}
   \usepackage[T2A]{fontenc}
   \usepackage{epstopdf}

\begin{document}
\title{Evolution of sub-spaces at high and low energies}

\author{\firstname{Arkady A.}~\surname{Popov}} \email{apopov@kpfu.ru}\affiliation{N.~I.~Lobachevsky Institute of Mathematics and Mechanics, Kazan  Federal  University, \\ 420008, Kremlevskaya  street  18,  Kazan,  Russia}
\author{\firstname{Sergey G.}~\surname{Rubin}}\email{sergeirubin@list.ru} \affiliation{National Research Nuclear University MEPhI (Moscow Engineering Physics Institute),\\ 115409, Kashirskoe shosse 31, Moscow, Russia}
\affiliation{N.~I.~Lobachevsky Institute of Mathematics and Mechanics, Kazan  Federal  University, \\ 420008,
Kremlevskaya  street  18,  Kazan,  Russia}

\begin{abstract}
The evolution of sub-spaces in the framework of gravity with higher derivatives is studied. Numerical solutions to exact differential equations are found. It is shown that the initial conditions play crucial role in the space dynamic. Appropriate metrics describing an expanding and a stationary sub-space shed light on the well-known question: why our 3-dim space is large but an extra space is small and stable (if exists)? It is assumed that the values of parameters at high energies strongly depend on uncontrolled quantum corrections and, hence, are not equal to their values at low energies.  Therefore, there is no way to trace solutions throughout the energy range, and we restrict ourselves to the sub-Planckian and the inflationary energies.
\end{abstract}
\maketitle
%

\section{Introduction}

The origin of our Universe remains as an unresolved problem up to now. It is usually assumed that its nucleation is related to the quantum processes at high energies \cite{1994CQGra..11.2483V,Bousso:1998ed,2009arXiv0909.2566H,Yurov:2005zn}.
The probability of its creation remains unclear in spite of wide discussion, see e.g. \cite{Kubyshin:1989mj,Firouzjahi:2004mx,2013arXiv1311.0220C}. Here we are interested in the subsequent classical evolution of the metrics rather than a calculation of this probability. It is assumed that manifolds can be described by specific metrics after their nucleation. After nucleation, these manifolds evolve classically forming a set of asymptotic manifolds, one of which could be our Universe.

The complexity of the problem is greatly aggravated by two factors. First, the metric evolution should lead to the formation of our Universe with the strong fine-tuning of the observational parameters \cite{2007unmu.book..231D,Bauer:2010wj}. Second,
the inclusion of extra dimensions is of particular interest because the idea of extra space is widely used in modern research. They shed light to such issues as the grand unification \cite{ArkaniHamed:1998rs,1999NuPhB.537...47D}, neutrino mass \cite{2002PhRvD..65b4032A}, the cosmological constant problem \cite{2001PhRvD..63j3511S,2003PhRvD..68d4010G,2016arXiv160907361R} and so on. In this regard, the immediately aroused question is: why specific number of dimensions are asymptotically compact and stable while others expand \cite{2007JHEP...11..096G,2002PhRvD..66b4036C,2002PhRvD..66d5029N}? Which specific property of subspace leads to its quick growth? Sometimes one of the subspaces is assumed to be FRW space by definition \cite{2013bhce.conf.....B}. There are many attempts to clarify the problem, mostly related to introduction of fields other than gravity. It may be a scalar field \cite{2017PhRvD..95j3507K,2007JHEP...11..096G} (most used case) and gauge fields \cite{2009PhRvD..80f6004K} for example. A static solution can be obtained using the Casimir effect \cite{1984NuPhB.237..397C,2018GrCo...24..154B} or form fields  \cite{1980PhLB...97..233F}. Another possibility was discussed in \cite{1984PhRvD..30..344Y,2013GReGr..45.2509B}: it was shown that if the scale factor $a(t)$ of our 3D space is much larger than the growing scale factor $b(t)$ of the extra dimensions, a contradiction with observations can be avoided.

In our previous article \cite{Lyakhova:2018zsr} we studied evolution of manifolds after their creation on the basis of pure gravitational Lagrangian with higher derivatives. It was shown analytically and confirmed numerically that an asymptotic growth of the manifolds depends weakly on initial conditions. We have shown that the initial conditions can be a reason of nontrivial solutions (funnels) and studied their properties. A number of final states of metric describing our Universe is quite poor if we limit ourselves with a maximally symmetric extra space and the $f(R)$ gravity.

In this article, we continue to study the Universe evolution at the sub-Planckian scale. The space $V_D$ is assumed to be the direct product $T\times V_3\times W_3$ of the time and the two maximally symmetric manifolds with positive curvature. Both sub-spaces are born with the size of the order of the Planck scale or more.
Most of the resulting sub-spaces are characterized by initial metrics, which lead to the growth of both 3-dimensional sub-spaces, which clearly contradicts the observations. We have found a set of metrics that could lead to the observable space-time metric.

The action used should not contradict the observations at the low and intermediate energies. More definitely, the Lagrangian parameters at low energies should be chosen in such a way to supply (almost) Minkowski space for the modern Universe, the stationarity of the extra space metric at the modern epoch, and reproduce the inflationary stage with the Hubble parameter of the order of $H\sim 10^{13}$GeV. On the contrary, the parameter values at the sub-Planckian energies are free from such restrictions. Indeed, the quantum corrections to the parameter values cause their dependence on the energy scale. There is a lot of literature devoted to this subject, see e.g. \cite{Babic:2001vv}. The quantum corrections to the physical parameters are large at the very high energies where interactions of all kinds of fields must be taken into account. Our knowledge of the physical parameters at the low energies is blurring at sub-Planckian scale.

The gravity with higher derivatives is widely used in modern research despite the internal problems inherent in this approach \cite{2015arXiv150602210W}.  Most promising model of inflation is the Starobinsky model based on pure gravitational action. Attempts to avoid the Ostrogradsky instabilities are made \cite{2017PhRvD..96d4035P} and extensions of the Einstein-Hilbert action attract much attention. Promising branch of such models is based on the Gauss-Bonnet Lagrangian \cite{2010IJGMM..07..797I,2018EPJC...78..373P} and its generalization to the Horndeski gravity \cite{Starobinsky:2016kua}. These models were adjusted to obtain differential equations of the second order so that the Ostrogradsky theorem is not dangerous for them. Our model contains $f(R)$ gravity with the Gauss-Bonnet term acting in $D=7$ dimensions.
The similar research was performed in \cite{Shafi:1986vv}.
The authors considered the action containing all scalars made from the Riemann tensor up to the second term multiplied by an arbitrary factor. They discuss asymptotic evolution of the two subspaces with one of them tending to a stationary state. The equations were simplified by the slow motion approximation. Our analysis is based on exact solutions to the nonlinear equations and hence can be applied to the whole variety of initial metrics. This ability is important for our study, which aims to study the role of the initial conditions.

Throughout this paper we use the conventions for the curvature tensor $R_{ABC}^D=\partial_C\Gamma_{AB}^D-\partial_B\Gamma_{AC}^D+\Gamma_{EC}^D\Gamma_{BA}^E-\Gamma_{EB}^D\Gamma_{AC}^E$ and for the Ricci tensor $R_{MN}=R^F_{MFN}$. The units $\hbar = c = 1$ are also used.

\section{The model}\label{eq}

The action
\begin{eqnarray}\label{Seff}
&&S_{eff}=\frac12 m_D^{D-2}\int d^D x \sqrt{|g_D|}\Big[f(R)+ c_1R_{AB}R^{AB} + c_2R_{ABCD}R^{ABCD}\Big]
\end{eqnarray}
can be considered as a basis of an effective theory \cite{2007ARNPS..57..329B}. Here $f(R)$ is an arbitrary function of the Ricci scalar $R$, $m_D$ is the D-dimensional Planck mass and $c_1,c_2$ are parameters of the Lagrangian. Such action was used in \cite{Shafi:1986vv} to study the extra space dynamic in the slow motion limit. Its particular case, the Gauss-Bonnet Lagrangian
\begin{eqnarray}
\label{L_GB}
{\cal L}_{GB} = k\sqrt{-g}\biggr\{R^2 - 4R_{AB}R^ {AB} + R_{ABCD}R^{ABCD}\biggl\}
\end{eqnarray}
is the appropriate starting point because of the absence of higher derivatives in the equations of motion. $f(R)$ extensions of the gravitational action is also applied in this content \cite{Troisi:2017tuw}.

The action
\begin{equation}\label{Sgen}
S_{gen}=\frac12 m_D^{D-2}\int d^D x \sqrt{|g_D|}\Big[\tilde{f}(R)+k\biggr\{R^2 - 4R_{AB}R^{AB} + R_{ABCD}R^{ABCD}\biggl\}\Big],
\end{equation}
containing the Gauss-Bonnet term plus $\tilde{f}(R)$ term (a function of the Ricci scalar) will be used throughout the paper. Such action was used to describe the dark energy phenomenon \cite{2007PhLB..651..224N} for example. The action \eqref{Sgen} is the particular case of the action \eqref{Seff} provided that
 $\tilde{f}(R)=f(R) - kR^2$ and $c_1=-4k, c_2 =k$. In what follows we restrict ourselves to the quadratic function
 \begin{equation}
 f(R)=  aR^2+bR+c
 \end{equation}
 ($b=1$ without the loss of generality).
 
 A separate problem is the values of the physical parameters. It is well known that values of Lagrangian parameters depend on the energy scale. There are at least two approaches to calculate this dependence - the Renormalization group analysis \cite{Babic:2001vv} and the Effective field theory \cite{2007ARNPS..57..329B}.
In any case, the physical parameters are functions $a(M),k(M),c(M)$ of an energy scale $M$. The next Section relates to the highest energy scale of the D-dim Planck mass, $M\simeq m_D$ and hence we may find the specific values $a(m_D),k(m_D),c(m_D)$ but not functions $a(M),k(M),c(M)$.
Unfortunately, there are no way to connect accurately these functions at the inflationary ($M_{infl}$) scale and at the sub-Planckian scale. Some restrictions to the parameters $a(M_{infl}),k(M_{infl}),c(M_{infl})$  are found in Section \ref{low}.

\section{Subspaces evolution at the Planck energies}\label{planck}

In this section, the evolution of extra spaces at highest energies $m_D$ is analyzed on the basis of action \eqref{Sgen}. Both extra spaces are described by 3-dim maximally symmetrical metrics with positive curvature
\begin{eqnarray}\label{metric2}
ds^2 = &&dt^2-e^{2\alpha(t)} m_D^{-2} [dx^2 +\sin^2(x)dy^2 +\sin^2(x)\sin^2(y)dz^2]
\nonumber\\&& \quad
-e^{2\beta(t)} m_D^{-2} [d\theta^2 + \sin^2(\theta)\, d\phi^2+ \sin^2(\theta) \sin^2(\phi)\, d\psi^2].
\end{eqnarray}
Below, we show that one of the sub-spaces can evolve into a space with a large volume while the second one - into a small static extra space.

Einstein's equations for this model are
	\begin{eqnarray}\label{AB}
	&&-\frac{1}{2}\tilde{f}(R)\delta_B^A + (R_B^A +\nabla^{A}\nabla_{B} - \delta_B^A \square) \tilde{f}_R
    +k\Big[ -8 {R^{AC}}_{;BC}  -12R^{AC} R_{CB}
    \nonumber \\ &&
     +2\delta^A_B R_{CD} R^{CD}
    -\frac{\delta^A_B}{2}  R^{CDEF} R_{CDEF} +2 R^{ACDE} R_{BCDE}  +4 {R^{;A}}_{;B}
    \nonumber \\ &&
     +4 {R^{CAD}}_B^{\ \cdot} \ R_{CD} -\frac{\delta^A_B}{2} R^2 +2R R^A_B \Big] = 0,
	\end{eqnarray}

Our aim is to study the behaviour of the two scale factors at highest energies $E \lesssim m_D$ by solving the main system \eqref{tt2}, \eqref{xx2}, \eqref{thth2}, \eqref{R}. The main goal is to find a solution - $\alpha(t)=Ht, \ H=const$ and $\beta(t)=const$. The Hubble parameter $H$ at high energies is not related to those at the inflationary stage.

The nontrivial system of equations \eqref{AB} are
	\begin{eqnarray}\label{tt2}
	&& -36 k \left[ {\dot{\alpha}}^3 \dot{\beta} +3 {\dot{\alpha}}^2 {\dot{\beta}}^2
    + \dot{\alpha} {\dot{\beta}}^3
    +e^{-2 \alpha} \dot{\beta} \left( \dot{\alpha} + \dot{\beta} \right)
    +e^{-2 \beta} \dot{\alpha} \left( \dot{\alpha} + \dot{\beta} \right)
    +e^{-2 \alpha} e^{-2 \beta}
    \right]
    \nonumber \\ &&
    -3 \tilde{f}_{RR} \dot{R} \left( \dot{\alpha} +\dot{\beta} \right)
    +3 \tilde{f}_R \left( \ddot{\alpha} + \ddot{\beta} +{\dot{\alpha}}^2 +{\dot{\beta}}^2 \right) -\frac12 \tilde{f}(R)=0,
	\end{eqnarray}
	\begin{eqnarray}\label{xx2}
	&& -12 k \left\{ 2 \ddot{\alpha} \dot{\beta}\left( \dot{\alpha} + \dot{\beta} \right)
    +\ddot{\beta} \left( {\dot{\alpha}}^2 +4 \dot{\alpha}\dot{\beta} +{\dot{\beta}}^2 \right)
    +2 {\dot{\alpha}}^3 \dot{\beta} +6 {\dot{\alpha}}^2 {\dot{\beta}}^2 + 6 \dot{\alpha} {\dot{\beta}}^3 +{\dot{\beta}}^4
    \right. \nonumber \\ &&  \left.
    +e^{-2 \alpha} \left[ \ddot{\beta} +2{\dot{\beta}}^2 \right]
    +e^{-2 \beta} \left[ 2 \ddot{\alpha} +\ddot{\beta} + 3{\dot{\alpha}}^2 +2 \dot{\alpha}\dot{\beta} +{\dot{\beta}}^2 \right] +e^{-2 \alpha} e^{-2 \beta}
    \right\}
    - \tilde{f}_{RRR} {\dot{R}}^2
    \nonumber \\ &&
    - \tilde{f}_{RR} \left[\ddot{R} +\dot{R} \left(2 \dot{\alpha} +3\dot{\beta} \right) \right]
    + \tilde{f}_R \left( \ddot{\alpha} +3{ \dot{\alpha}}^2 + 3 { \dot{\alpha}} \dot{\beta} +2 e^{-2 \alpha} \right) -\frac12 \tilde{f}(R)=0,
	\end{eqnarray}
	\begin{eqnarray}\label{thth2}
	&& -12 k \left\{ \ddot{\alpha} \left( {\dot{\alpha}}^2  +4\dot{\alpha} \dot{\beta} + {\dot{\beta}}^2 \right) +2 \ddot{\beta} \dot{\alpha} \left( \dot{\alpha}+ \dot{\beta} \right)
    +{\dot{\alpha}}^4 +6 {\dot{\alpha}}^3 \dot{\beta} +6 {\dot{\alpha}}^2 {\dot{\beta}}^2 + 2 \dot{\alpha} {\dot{\beta}}^3
    \right. \nonumber \\ &&  \left.
    +e^{-2 \alpha} \left[ \ddot{\alpha} +2\ddot{\beta}+{\dot{\alpha}}^2 +2 \dot{\alpha}\dot{\beta} +3{\dot{\beta}}^2 \right]
    +e^{-2 \beta} \left[ \ddot{\alpha} +2{\dot{\alpha}}^2 \right] +e^{-2 \alpha} e^{-2 \beta}
    \right\}
    - \tilde{f}_{RRR} {\dot{R}}^2
    \nonumber \\ &&
    - \tilde{f}_{RR} \left[\ddot{R} +\dot{R} \left(3 \dot{\alpha} +2\dot{\beta} \right) \right]
    + \tilde{f}_R \left( \ddot{\beta} + 3 { \dot{\alpha}} \dot{\beta} +3{ \dot{\beta}}^2  +2 e^{-2 \beta} \right) -\frac12 \tilde{f}(R)=0,
	\end{eqnarray}
where we have kept in mind denotations $\partial_t \tilde{f}_R=\tilde{f}_{RR}\dot{R}$ and $\partial^2_t \tilde{f}_R=\tilde{f}_{RRR}\dot{R}^2 + \tilde{f}_{RR}\ddot{R}$.
The Ricci scalar is
	\begin{eqnarray}\label{R}
	&& R= 6 \left(\ddot{\alpha} +  \ddot{\beta} + 2{\dot{\alpha}}^2 + 3 \dot{\alpha} \dot{\beta} +2 {\dot{\beta}}^2 + e^{-2 \alpha} + e^{-2 \beta} \right).
	\end{eqnarray}
Here and in the following the units $m_D=1$ are assumed.
	
For calculations, it is convenient to consider the Ricci scalar $R(t)$ as the additional unknown function and interpret definition \eqref{R} as the fourth equation.
Three equations of this system (for example, \eqref{xx2}, \eqref{thth2}, \eqref{R})
can be solved with respect to the higher derivatives $\ddot{\alpha}, \ddot{\beta}, \ddot{R}$.
Then, substitution $\ddot{\alpha}$ and $\ddot{\beta}$ into equation (\ref{tt2}) gives equation
\begin{eqnarray} \label{eq002}
&& -36 k \left[ {\dot{\alpha}}^3 \dot{\beta} +3{\dot{\alpha}}^2 {\dot{\beta}}^2 +{\dot{\alpha}} {\dot{\beta}}^3
+e^{-2\alpha}\dot{\beta}\left(\dot{\alpha} +\dot{\beta} \right) +e^{-2\beta}\dot{\alpha}\left(\dot{\alpha} +\dot{\beta} \right)
+e^{-2\alpha} e^{-2\beta}\right]
\nonumber \\ &&
-3\left( \dot{\alpha} +\dot{\beta} \right)\dot{R} \tilde{f}_{RR}
+\left(-3{\dot{\alpha}}^2 -9\dot{\alpha}\dot{\beta} -3\dot{\beta} -3e^{-2\alpha} -3e^{-2\beta} +\frac{R}{2}  \right)\tilde{f}_R -\frac{\tilde{f}}{2}
=0,
\end{eqnarray}
which plays the role of restriction to the solutions of the coupled second order differential equations.
This can be checked, for example, by writing the set of four equations \eqref{xx2}, \eqref{thth2}, \eqref{R}, \eqref{eq002}
as an equivalent set of (six) coupled first-order equations plus one algebraic equation.
The equation \eqref{eq002} reduces to the algebraic transcendental equation, i.e., it is a constraint.
The complete set of initial conditions may be chosen in the form
$\alpha (t_0), \beta (t_0), R(t_0), \dot{\alpha} (t_0), \dot{\beta} (t_0)$, and $\dot{R}(t_0)$. These initial conditions are not independent due to equation \eqref{eq002}. The latter will be used to derive an exact relation between these initial data.

Natural values of the parameters are assumed to be of the order of the Planck scale:  $a\sim k\sim m_D^{-2}, c\sim  m_D^{2} $. They are not related to the observational values because of strong and uncontrolled contribution of the quantum corrections at sub-planckian energies. Hence, they are considered as free parameters.

Our analysis revealed a complex dynamic of the sub-spaces depending on the values of the parameters $k, a, c$ and the initial metric. There are several variants for the metric evolution:

i) both sub-spaces expand at equal rates;

ii) both sub-spaces expand at different rates;

iii) one of the sub-spaces expands while the other remains constant.

iv) one or both sub-spaces shrink. Looking ahead, we note that such solutions do exist (see solid line in the left panel of Fig. \ref{figb}). This means that there exists such a set of initial metrics for which manifolds just nucleated, come back to the space-time foam. The destiny of manifold depends on its initial metric. Our main aim is to find a set of those metrics that could correspond to the evolution of our Universe and hence validate the issue iii)

Let us find the asymptotic solution of the expanding sub-spaces in the form
\begin{equation} \label{asHH}
	\alpha(t) = H_1 t,\quad 	\beta(t)=H_2 t,\quad H_1 > 0, H_2 > 0,\quad t\rightarrow\infty.
\end{equation}
In this case we can strongly simplify the equations of motion \eqref{tt2}, \eqref{xx2}, \eqref{thth2} which are transformed into the system
\begin{eqnarray} \label{asHHtt}
&&- H_1^2 -3 H_1 H_2 - H_2^2 +\Big(24 H_1^3 H_2+18 H_1^2 H_2^2+24 H_1 H_2^3\Big) k
\nonumber \\
&& -\Big(36 H_1^3 H_2 +54 H_1^2 H_2^2 +36 H_1 H_2^3\Big) a -\frac{c}{6}=0,
\end{eqnarray}
\begin{eqnarray} \label{asHHxx}
&&- H_1^2  -2 H_1 H_2 - 2 H_2^2 +\Big(4 H_1^3 H_2+18 H_1^2 H_2^2+24 H_1 H_2^3 +20 H_2^4\Big) k
\nonumber \\
&&-\Big(12 H_1^3 H_2 +42 H_1^2 H_2^2 +48 H_1 H_2^3 +24 H_2^4 \Big) a -\frac{c}{6}=0,
\end{eqnarray}
\begin{eqnarray} \label{asHHthth}
&&- 2 H_1^2 -2 H_1 H_2 - H_2^2 +\Big(20 H_1^4 +24 H_1^3 H_2+18 H_1^2 H_2^2+4 H_1 H_2^3\Big) k
\nonumber \\
&& -\Big(24 H_1^4 +48 H_1^3 H_2 +42 H_1^2 H_2^2 +12 H_1 H_2^3\Big) a -\frac{c}{6}=0.
\end{eqnarray}
Only two of the three equations (\ref{asHHtt}, \ref{asHHxx}, \ref{asHHthth}) are independent (the combination $H_1 \cdot$ Eq.\eqref{asHHxx} +$H_2 \cdot$  Eq.\eqref{asHHthth} $-(H_1+H_2)  \cdot$ Eq.\eqref{asHHtt} = 0 is an identity).
These equations have two different solutions. The simplest one is characterized by equal asymptotes.  In this case, equations \eqref{asHHtt}, \eqref{asHHxx} and \eqref{asHHthth} are reduced to a single equation
\begin{equation} \label{HH1}
(66 k -126 a) H^4 -5H^2 -\frac{c}{6}=0,
\end{equation}
with the solution
\begin{equation} \label{HH11}
H=H_1=H_2=\sqrt{\frac{5 \pm\sqrt{25 +44 k c -84 a c}}{12(11 k -21 a)}}.
\end{equation}

\begin{figure}[ht!]
\includegraphics[width=7cm]{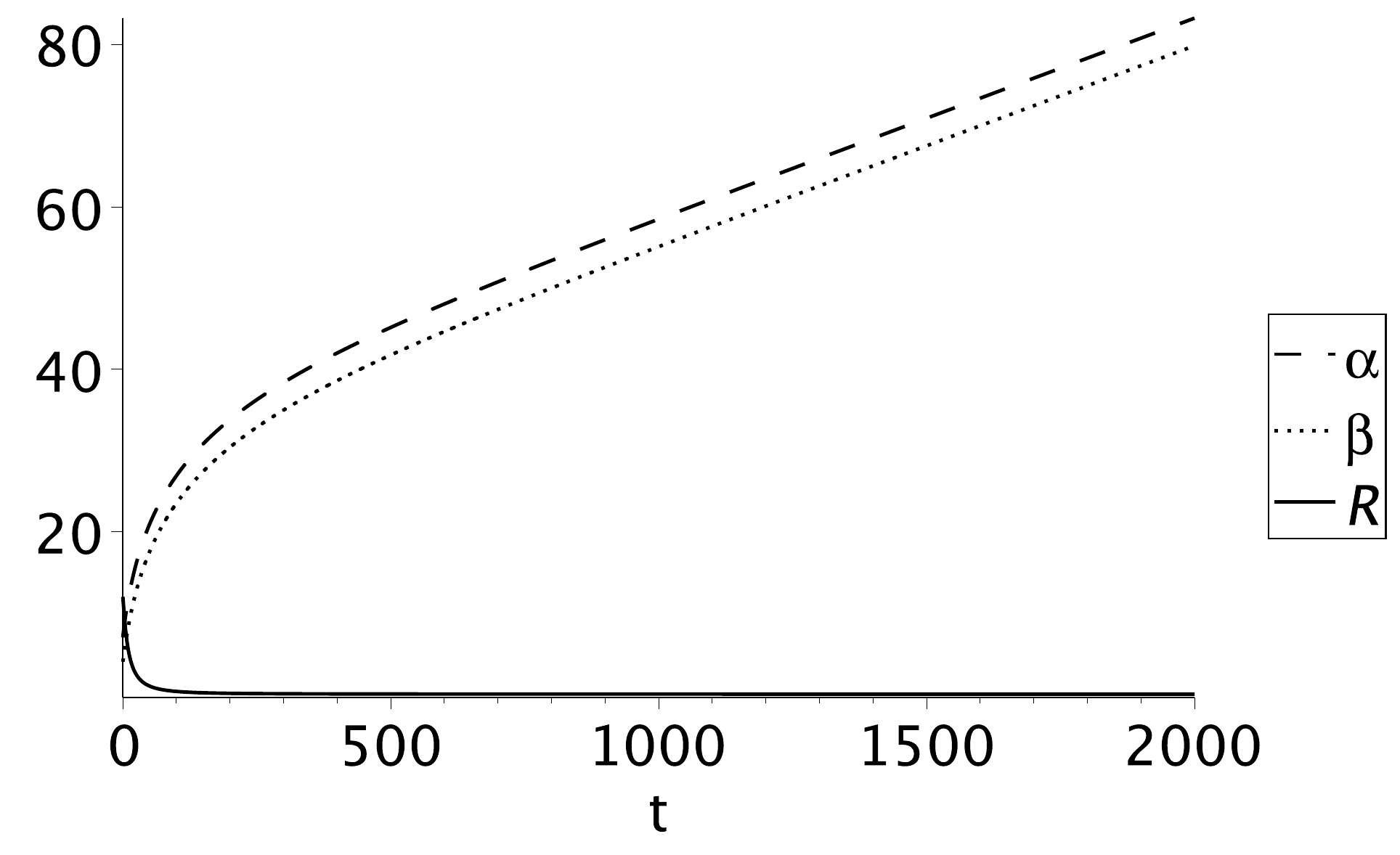} \quad \includegraphics[width=7cm]{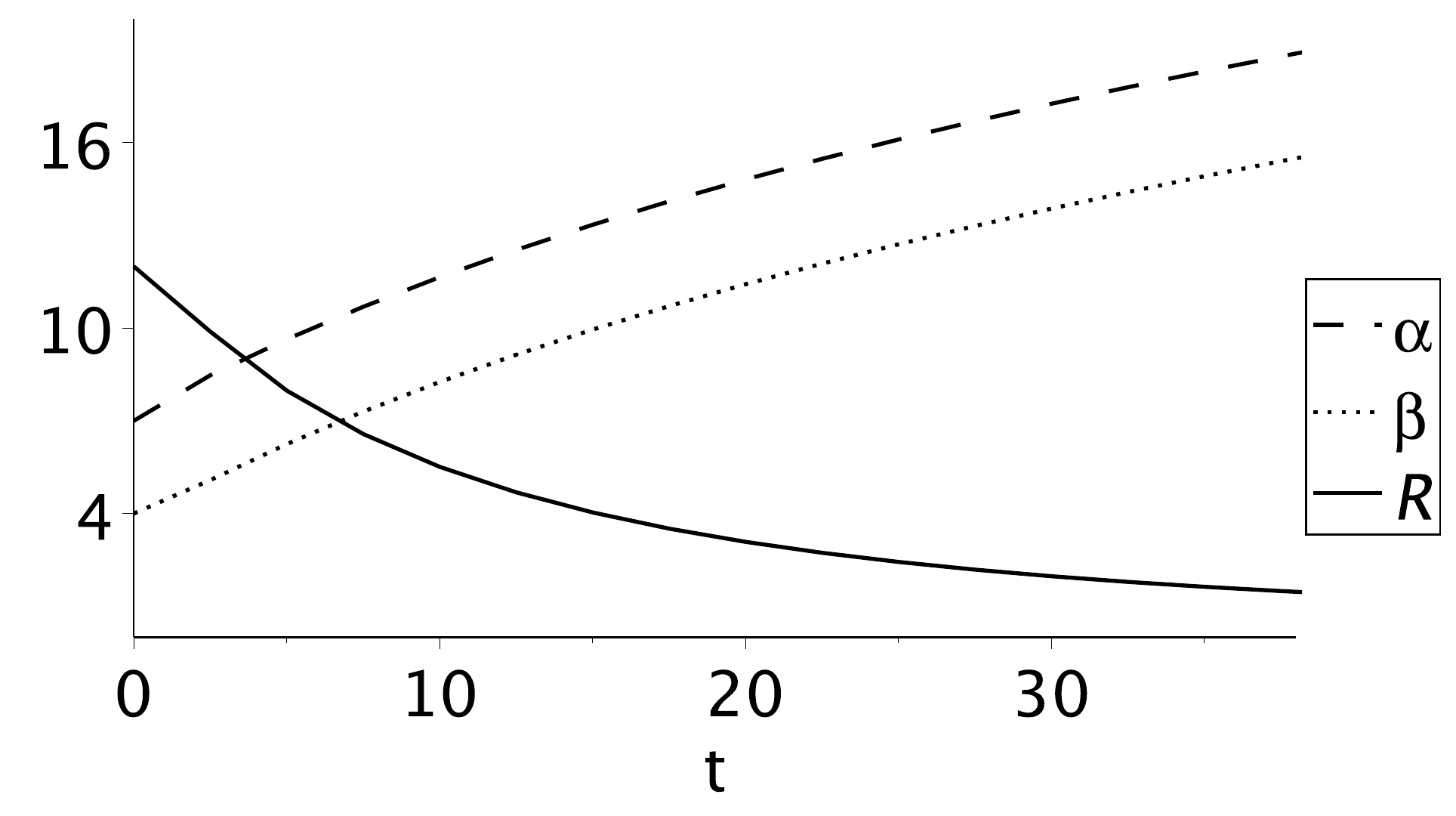}
\caption{Numerical solution to the system of equations \eqref{xx2}, \eqref{thth2} and \eqref{R} for initial conditions $ \alpha(0)=7, \ \beta(0)=4, \ \dot{\alpha}(0)=1, \ \dot{\beta}(0)=0, \ \dot{R}(0)=0$. The initial condition $R(0)\simeq 11.99901$ is found from equation \eqref{eq002}. The  values of  $ H_1 =H_2 \simeq 0.02461$ coincide with those found from expression  (\ref{HH11}).  The Lagrangian parameters are  $k = 500, a = 200, c = -0.001$.}
 \label{fig1}
\end{figure}

In this paper, we are interested in the dependence of solutions to system \eqref{tt2}, \eqref{xx2}, \eqref{thth2} on the initial conditions. In this particular case, the initial conditions leading to asymptotic behaviour \eqref{HH11} are found. An example of solution with such asymptotes is represented in FIG.~\ref{fig1}.

\begin{figure}[ht!]
\includegraphics[width=7cm]{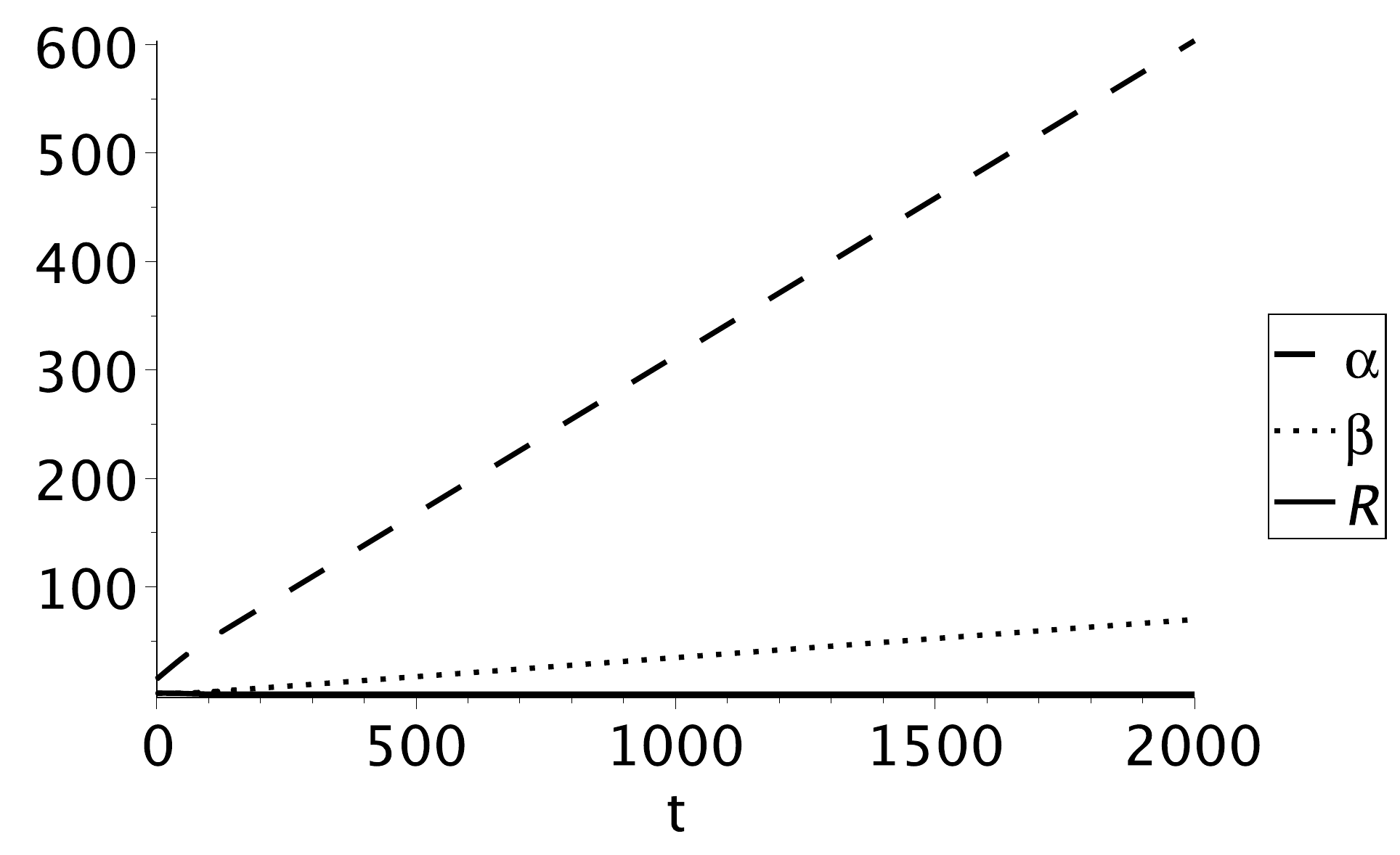} \quad \includegraphics[width=7cm]{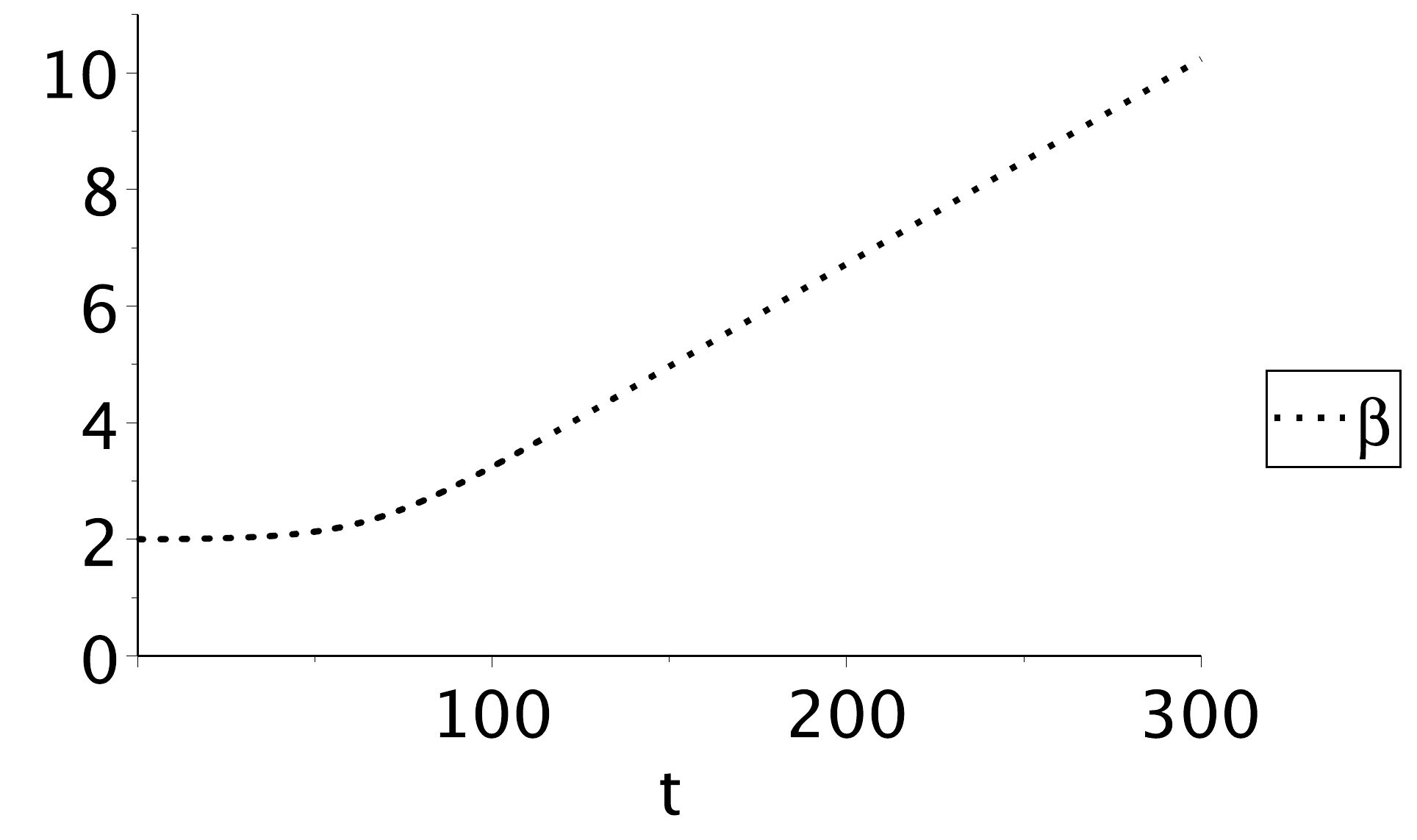}
\caption{Numerical solution to the system of equations \eqref{xx2}, \eqref{thth2} and \eqref{R} for initial conditions $ \alpha(0)=15, \ \beta(0)=2, \ \dot{\alpha}(0)\simeq 0.4046668, \ \dot{\beta}(0)=0, \ \dot{R}(0)=0$. $R(0)\simeq 2.0912618$ is found from equation \eqref{eq002}. The  values of $H_1\simeq 0.290696, H_2 \simeq 0.035163$ coincide with those found from \eqref{asHHtt}, \eqref{asHHxx}, \eqref{asHHthth} equations. The Lagrangian parameters are $k = -2.98, a = -2.77, c = -0.49$.}
\label{fig2_}
\end{figure}

The choice of another set of physical parameters can change the picture. Sub-spaces grow exponentially at different rates ($H_1 \neq H_2$) as is represented in FIG.~\ref{fig2_}. In this case, as well as in the previous one, both sub-spaces appear to be large at the present time and hence such growth is inadmissible.

\begin{figure}[ht!]
\includegraphics[width=10cm]{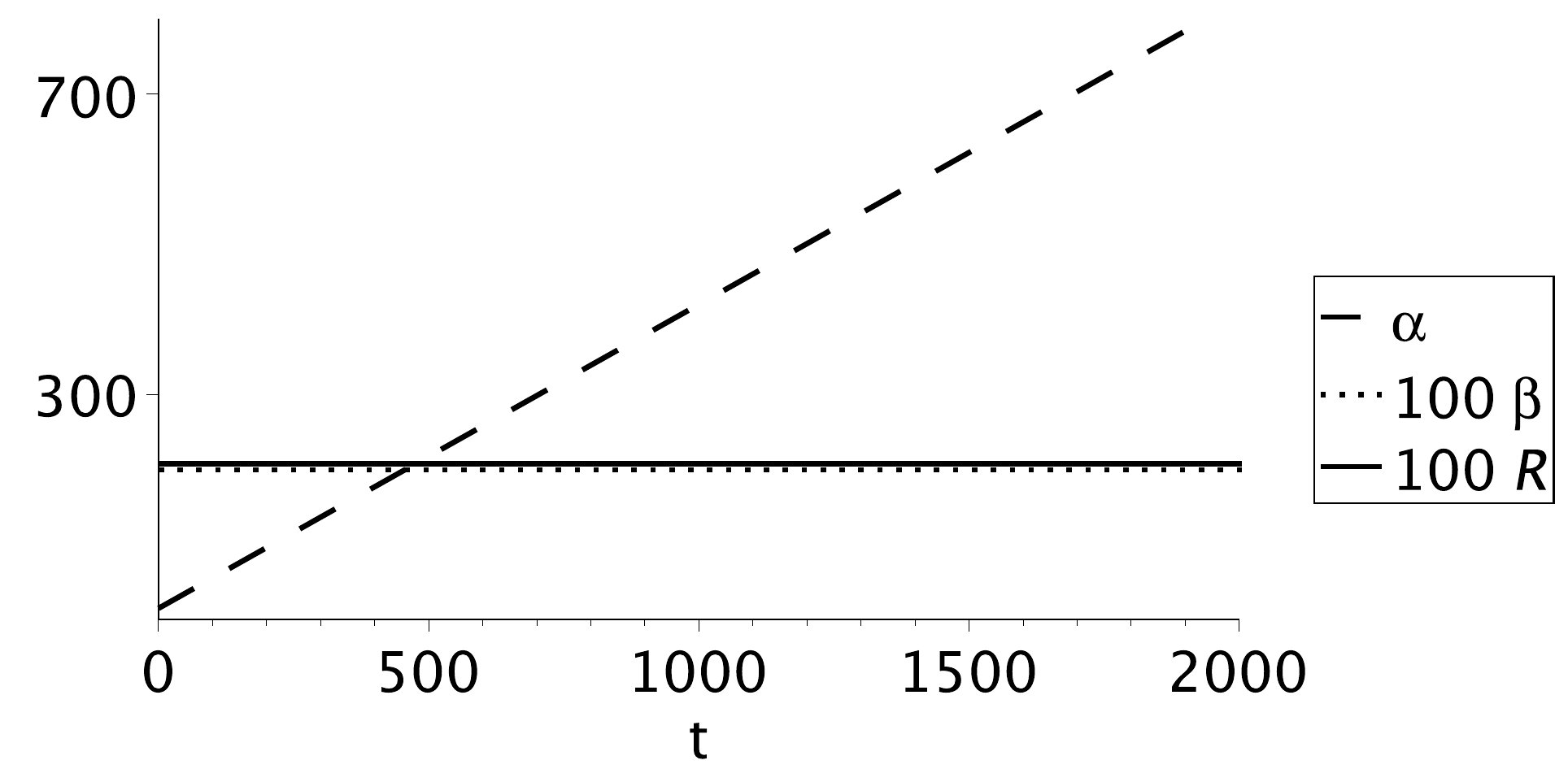} 
\caption{Numerical solution to the system of equations \eqref{xx2}, \eqref{thth2} and \eqref{R} for initial conditions $ \alpha(0)=15, \ \beta(0)=b_c \simeq 1.9930281, \ \dot{\alpha}(0)=H \simeq 0.4046668 $ ($b_c, H$ are found from equations \eqref{ttH_}, \eqref{ththH_}) \ $\dot{\beta}(0)=0, \ \dot{R}(0)=0$. $R(0)\simeq 2.0764993$ is found from equation \eqref{eq002}. For the found numerical solution $k = -2.98, a = -2.77, c = -0.49$.}
\label{oneexpon_}
\end{figure}

According to the observations, the most promising case is iii) - one of the sub-spaces expands while the other remains constant:
\begin{equation} \label{asHb0}
	\alpha(t) = H t,\quad 	\beta(t)=\beta_{as},\quad H>0, \quad \beta_{as}>0, \quad t\rightarrow\infty .
\end{equation}
Asymptotic regime \eqref{asHb0} can be obtained by substituting these expressions into equations \eqref{tt2}, \eqref{xx2}, \eqref{thth2}. This leads to the algebraic system
	\begin{eqnarray}
&&    -\left( 1+\frac{12 a}{e^{2 \beta_{as}}} \right) H^2 +\frac{6(k-a)}{e^{4 \beta_{as}}} -\frac{1}{e^{2 \beta_{as}}} -\frac{c}{6}=0, \label{ttH_}\\
&&	\left( 60 k -72 a \right) H^4 - \left(6+ \frac{24 a}{e^{2 \beta_{as}}} \right) H^2
    -\frac{6(k-a)}{e^{4 \beta_{as}}} -\frac{1}{e^{2 \beta_{as}}} -\frac{c}{2}=0\label{ththH_}
	\end{eqnarray}
which is used for the determination of the Hubble parameter $H$ and the asymptotic size $e^{\beta_{as}}$.

The results of numerical calculation of equations  \eqref{tt2}, \eqref{xx2}, \eqref{thth2} are represented in FIG. \ref{oneexpon_}. One of the sub-spaces grows exponentially while the second sub-space remains constant. The solution does not vary for a long time and looks stable. The results of numerical calculations with slightly different initial conditions $\beta (0)$ are shown in FIG. \ref{figb} (the first sub-space is growing permanently and is not represented in the figure). There are three types of metric behaviour depending on the initial value of $\beta(0)$  - after some time, the space is growing; the subspace tends to stay constant; the subspace shrinks and finally returns to the space-time foam (left panel).
\begin{figure}[ht!]
\includegraphics[width=7cm,height=3.5cm]{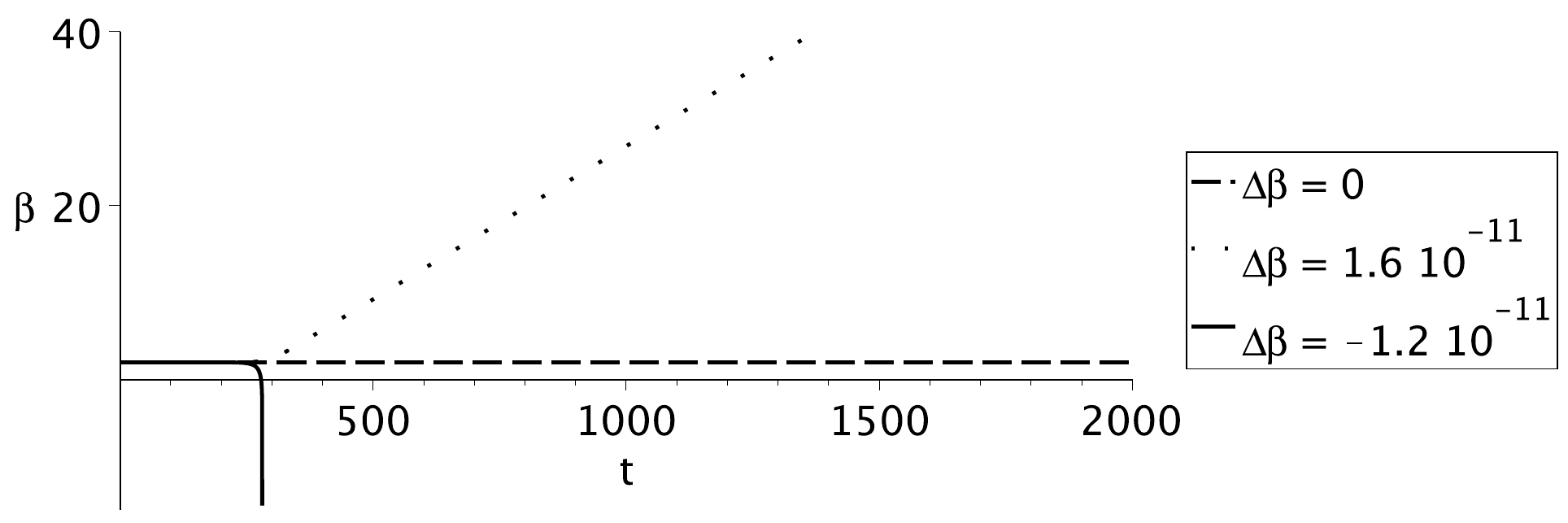} \quad \includegraphics[width=8.5cm,height=3.5cm]{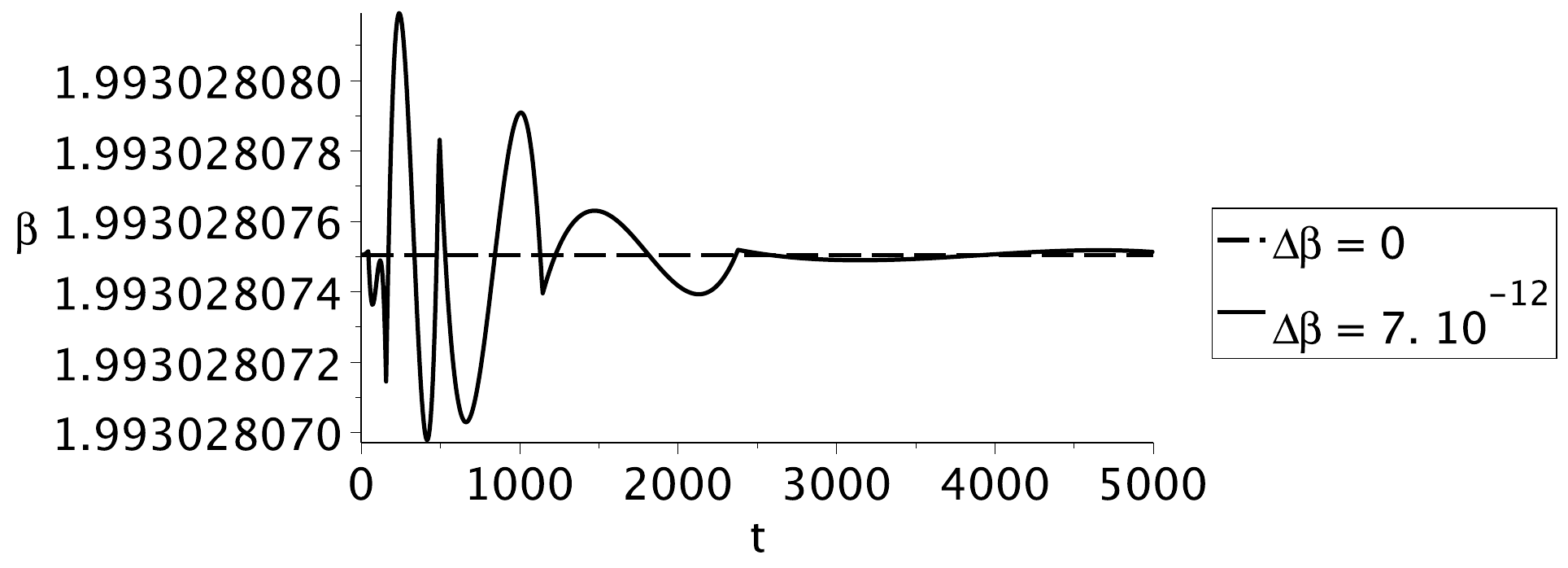}
\caption{Numerical solution to the system of equations \eqref{xx2}, \eqref{thth2} and \eqref{R} for initial conditions $ \alpha(0)=150, \ \beta(0) \simeq 1.9930281+\Delta \beta, \ \dot{\alpha}(0)\simeq 0.4046668, \ \dot{\beta}(0)=0, \ \dot{R}(0)=0$. $R(0)\simeq 2.0764993$ is found from equation \eqref{eq002}. The Lagrangian parameters are  $k = -2.98, a = -2.77, c = -0.49$. The numerical values of $H_1$ and $H_2$ coincide with those found from the (\ref{asHHtt}, \ref{asHHxx}, \ref{asHHthth}) or \eqref{ttH_}, \eqref{ththH_}) equations. Right panel illustrate motion in the stability region. }
\label{figb}
\end{figure}

\begin{figure}[ht!]
\includegraphics[width=10cm]{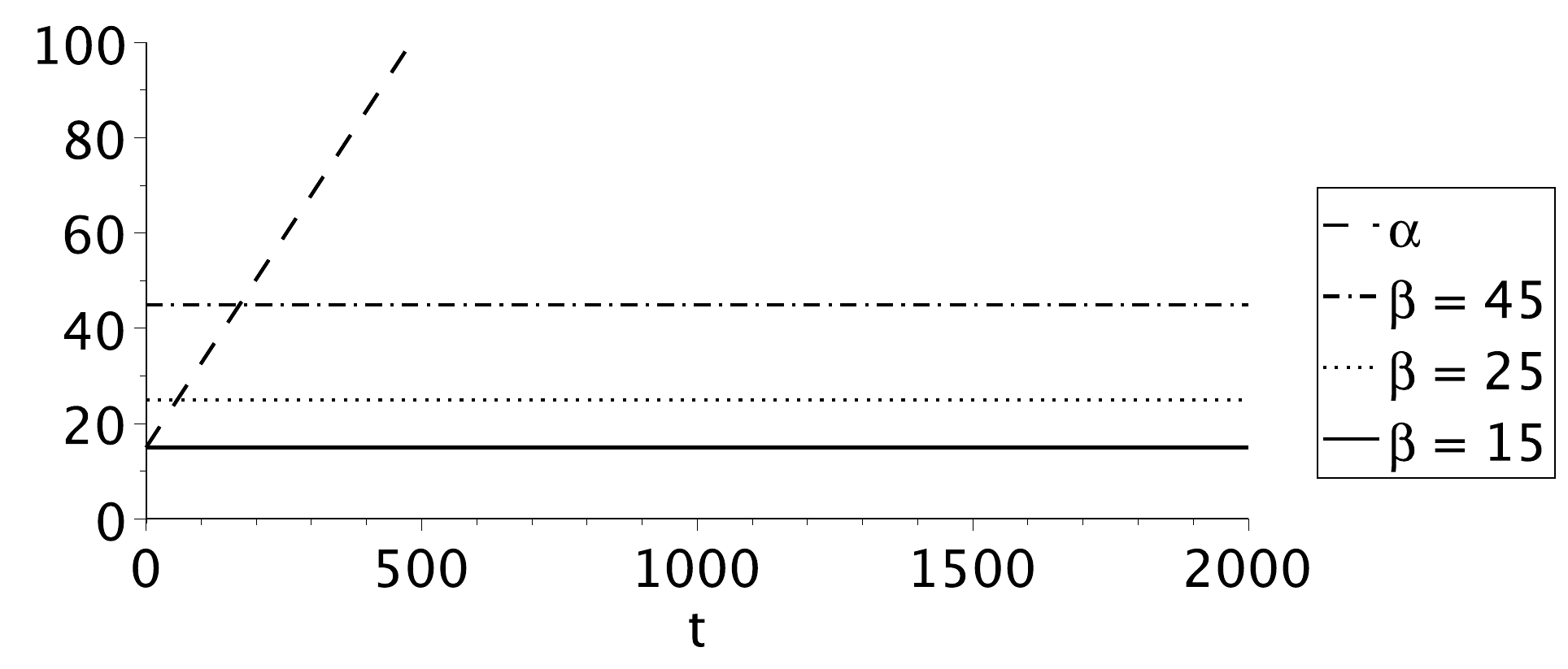}
\caption{The second extra space is constant for wide range of the initial metric $\beta(0)> 11$. The parameter values $k=-2, a=-3, c = -0.1875$. The initial values $\alpha(0)=15,  \ \dot{\alpha}(0)=H \simeq 0.17678, \ \dot{\beta}(0)=0, \ \dot{R}(0)=0.$ $R(0)\simeq 0.375$ is found from equation \eqref{eq002}.}
\label{constsolu}
\end{figure}

The stability region exists (left panel FIG.~\ref{figb}) but is very small, $\delta \beta \simeq 10^{-11}$ for the chosen set of parameters (right panel). The quantum fluctuations could easily push out the solution from the stability region and the second sub-space passes to the expanding or shrinking regime.
Therefore, it is worth seeking for another set of physical parameters with more appropriate stability conditions. The successful result is shown in FIG.~\ref{constsolu}. The size of one sub-space is expanded while the metric function of the second sub-space remains constant. Such solutions are realized in a broad interval of initial conditions.

In this section we have shown that the behavior of the extra spaces depends on initial conditions. We have found out promising conditions and the Lagrangian parameters for which one of the extra spaces remains small while the other expands quickly. We also have found initial conditions that lead to the expansion of both extra spaces. This case, unlike the previous one, contradicts observations.
The following sections are devoted to studying the model at the electro-weak and inflationary energy scales. In this case, we face the problem of choice of the Lagrangian parameters $a, k, c$ at the low (intermediate) energy scale, $M\lesssim 10^{13}$ GeV. Indeed, the quantum corrections to the parameter values are known to be extremely important at the Planck scale. Parameter values at high energies could differ drastically from those at low energies. It means that we can not use the values obtained earlier. Instead, we are able to obtain information from the observations. The metric of larger subspace is flat with great accuracy in the present time due to exceedingly small value of the dark energy density. Also, the Hubble parameter during the slow rolling inflation is more or less firmly established, $H\sim 10^{13}$GeV. This information is used below to impose restrictions to the  Lagrangian parameters.

It is assumed that the parameters $a(M),k(M),c(M)$ vary slowly for the scale $M$ smaller than the scale of inflation and we will neglect this dependence below.

\section{Subspaces evolution at intermediate and low energies}\label{low}

Previous study has shown that one can choose such parameter values and initial conditions for which one extra space exponentially grows and the other one remains constant starting from the sub-Planckian energies ($\sim 10^{18}$GeV). In this case, the beginning of the inflation is characterized by the two 3-dim sub-spaces - one of large and the other of extremely small size. The latter is treated as the extra space in the following.

The main goal of this section is to find those parameter values for which the extra space remains stable and its size is small. Restrictions to the parameters included in the Lagrangian \eqref{Sgen} can be imposed if we suppose that the sub-space $V_3$ is (almost) flat at present time due to extremely small value of the cosmological constant. Also, the knowledge of the Hubble parameter during the inflationary stage gives additional restriction which is discussed below.

\subsection{Parameter values at low energies}

There are several model parameters in action \eqref{Sgen} and some of them can be fixed by the low energy physics. It was shown in the previous Section that two subspaces $V_3$ and $W_3$ can evolve differently starting from the sub-Planckian scale of energy. It is assumed here that subspace $V_3$ acquires large volume and the subspace $W_3$ remains small up to the beginning of the inflation. This choice breaks the equivalence of these two sub-spaces and we may follow the method elaborated in \cite{Bronnikov:2005iz} to facilitate analysis.

Metric \eqref{metric2} of our 7-dim space leads to the Ricci scalar in the form
\begin{equation}\label{Tailor}
R=R_4 + R_3+ P_k; \quad P_k \equiv 6\ddot{\beta} +12 {\dot{\beta}}^2 +18\dot{\alpha} \dot{\beta}
\end{equation}
We assume that inequalities
\begin{equation}\label{ineq00}
P_k , R_4\ll R_{3}
\end{equation}
hold. This follows from the facts that the Ricci scalar $R_4$ of the sub-space $T\times V_3$ is small as compared to the Ricci scalar $R_3$ of the compact sub-space $W_3$ and the function $\beta (t)$ varies slowly during the inflation.

Expressions \eqref{Tailor} and \eqref{ineq00} validate the Tailor decomposition of the function $f(R)$ in action (\ref{Seff})
\begin{eqnarray}\label{actTailor}
&S=&\frac{v_{3}}{2}\int d^{4}x \sqrt{-g_4}e^{3\beta} [f(R_{3})+f'(R_{3})R_{4} + f'(R_{3})P_k  \nonumber \\
&& +c_1R_{AB}R^{AB}+c_2R_{ABCD}R ^{ABCD}+O(\epsilon^4)].
\end{eqnarray}
 Here $\epsilon^2 = R_4/R_3, v_{3}= 2\pi^2$ is the volume of $3$-dimensional sphere of the unit radius.

It is more familiar to work in the Einstein frame. To this end we have to perform conformal transformation
\begin{equation}\label{conform}
g_{ik}\rightarrow g_{ik}^{(E)}=|e^{3\beta} f'(R_3)|g_{ik};\quad R_4^{(E)}= \Big| e^{3\beta}f'(R_3)\Big|^{-1}R_4
\end{equation}
of the metric describing the subspace $M_{4}=T\times V_3$.
That leads to the action in the Einstein frame in the form \cite{Bronnikov:2005iz}
\begin{eqnarray}\label{Sscalar}
&& S_E=\frac{v_{3}}{2}  \int d^{4} x \sqrt{-g^{(E)}_4}\mbox{sign}(f')[R^{(E)}_{4} +  K_E(\phi)\dot{\phi}^{2} -2V_E (\phi) ], \\
&& K_E(\phi)=\frac{1}{2}\left(\frac{-3}{2\phi}+\frac{f''}{f'}\right)^2 +\left(\frac{f''}{f'}\right)^2+ \frac{3}{4\phi^{2}}+\frac{c_1+c_2}{f'\phi}, \label{KE}\\
&&V_E(\phi)=-\frac{\mbox{sign}(f')}{2|f'|^2} \left(\frac{ |\phi|}{6}\right) ^{\frac{3}{2}} \biggl[f(\phi)+\frac{c_1+c_2}{3}\phi^2\biggr],  \label{VE}
\end{eqnarray}
\begin{equation}\label{phi2}
\phi (t)\equiv R_{3}(t)= 6 e^{-2\beta(t)},
\end{equation}
where our physical intuition works properly.
In the modern epoch, the field $\beta$ is settled in a potential minimum.

The observable Planck mass is
\begin{equation}\label{MPl}
M_{P}^2=v_3 m_D^{D-2}=v_3=2\pi^2
\end{equation}
in the units $m_D=1$. The $D$-dim Planck mass is slightly smaller than the 4-dim Planck mass which doesn't contradict the limit $m_D> 10^{13}$GeV obtained in \cite{Nikulin:2019aka}.

The average metric of our Universe is the de Sitter metric with the extremely small Hubble parameter. This permits us to approximate the metric by the Minkowski one that strongly facilitate the analysis.
Some relations to the model parameters can be imposed in this case. The first condition
\begin{equation}\label{VMink}
V_E(\phi_0)=0;\quad V_E'(\phi_0)=0
\end{equation}
supplies the energy density of the Universe be zero. The inequalities
\begin{equation}\label{ineq}
V_E''(\phi_0)>0; \quad K_E(\phi_0)>0;
\end{equation}
are needed for the stability reasons. We also assume that the curvature of extra space is positive
\begin{equation}\label{ineq1}
\phi_0 >0.
\end{equation}
In the following, we put $b=1$ in $f(R)=aR^2 +bR +c$ without the loss of generality.
The algebraic equations \eqref{VMink} together with definition \eqref{VE} give a position of the potential minimum
\begin{equation}\label{cond1}
\phi_0 =\frac{-3}{2(3 a+ c_1+c_2)}>0
\end{equation}
and connection between the Lagrangian parameters valid for the Minkowski metric of the space $M_{4}=T\times V_3$
\begin{equation}\label{cond2}
c=\frac{3}{4(3a+ c_1+c_2)}.
\end{equation}
The first inequality in \eqref{ineq} gives
\begin{equation}\label{cond3}
  V_E''(\phi_0)=-\frac{|3a+c_1+c_2|\sqrt{-(3a+c_1+c_2)}}{24(c_1+c_2)|c_1+c_2|} > 0.
\end{equation}
The second inequality in \eqref{ineq} leads to the following expression:
\begin{equation}\label{cond4}
\frac{(3a +c_1 +c_2)^2}{6  (c_1 +c_2)^2} \Big(6a+c_1 +c_2 \Big)^2  >0
\end{equation}
and hence is true for any values of the parameters.

Finally, we have two inequalities
\begin{eqnarray}\label{cond5}
3 a +c_1+c_2<0,\quad   c_1+c_2<0
\end{eqnarray}
that lead to $f'(\phi_0)>0$ and formulas \eqref{cond1},  \eqref{cond2} which should be taken into account below.

We will continue to use the Gauss-Bonnet model, for which $c_1+c_2=-3k$,  that is,
\begin{eqnarray}\label{cond5_}
a -k<0,\quad   k > 0.
\end{eqnarray}
The parameter "$c$" is fixed by \eqref{cond2}, the Ricci scalar of the compact static extra space is known, see \eqref{cond1}  and \eqref{phi2}.

\subsection{ Moderate energies. Inflation}

We have obtained relations \eqref{cond5} between the parameters, connection \eqref{cond1} and expression \eqref{cond2} at the energy scale $M_{low}=M_{ew}\sim 10^2$GeV. Another restrictions can be obtained at the energy scale $M_{infl}\sim 10^{13}$GeV where the inflation takes place. We will assume that the parameters of our model varies slowly within the energy range $0-M_{infl}$ which is the widespread approach for inflationary models. This means that the restrictions obtained in the previous section hold at the inflation stage.

The idea is to find those parameter values that reduce potential \eqref{VE} to the form suitable for the inflationary scenario. The Ricci scalar $R_3$ plays the role of the inflaton $\phi$. The potential should have at least one minimum which is responsible for the (re-)heating after 50-60 e-folds during the inflation.

It is known that the inflation is finished when equality
\begin{equation}
    \epsilon=\frac{M_P^2}{16\pi}\left( \frac{V_E'}{K_E V_E}\right)^2 \sim 1,\quad \eta=\frac{M_P^2}{8\pi}\left( \frac{V_E'}{K_E}\right)' \frac{1}{V_E} \sim 1
\end{equation}
and
\begin{eqnarray}\label{H}
&&H^2 = \frac{8\pi V_E(\phi)}{3M_P^2}\sim 10^{-12}M_P^2  
\end{eqnarray}
are true \cite{Morris:2001ad}. The range of the Lagrangian parameters can be limited from approximate equalities
\begin{eqnarray}\label{cond6}
&&V_E(\phi_{end}) = \frac{3}{8\pi}H^2M_P^2 \sim \frac{3}{8\pi}10^{-12}M_P^4\simeq 5\cdot 10^{-12}, \label{H1} \\
&& \epsilon (\phi_{end}) \sim 1, \quad \eta (\phi_{end}) \sim 1 . \label{end}
\end{eqnarray}
Here, the expression \eqref{MPl} is used and
$\phi_{end}$ is the inflaton value at the end of the inflation.
We remind that the volume of the subspace $V_3$ is much greater than of the subspace $W_3$ at the beginning of the inflation.

\begin{figure}[ht!]
\includegraphics[width=10cm]{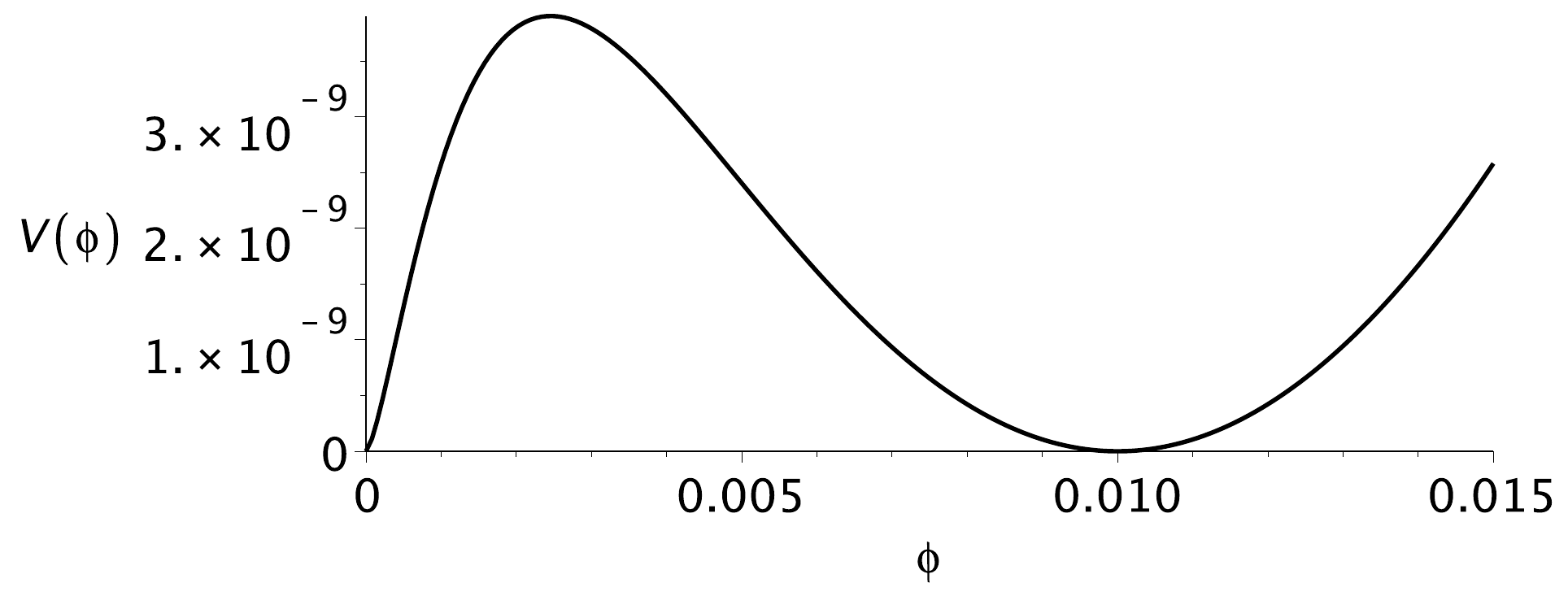}
\caption{ The potential \eqref{VE} for the parameter values $k(M_{low})=200, a(M_{low})=150$. The parameter "$c$" is fixed by \eqref{cond2}. }
\label{pot}
\end{figure}

The preliminary simulations indicate that appropriate form of the potential is realized for the parameter values $k(M_{low})=200, a(M_{low})=150$, see FIG. \ref{pot}.
The parameter $c$ is fixed by condition \eqref{cond2}.

The inflation is finished at  $\phi_{end}\simeq 0.0095$. This value was found by the solution of equation \eqref{end}. The present horizon arises when the field $\phi$ slightly differs from $\phi_{end}$. At the present time, the inflaton is placed in the potential minimum $$\phi_0 = 0.01.$$


One can conclude that the field moves near the potential minimum during the inflationary stage and after it.  Keeping in mind the connection \eqref{phi2} between the inflaton and the Ricci scalar of the compact extra space it can be concluded that this extra space is stable.

What about inequality \eqref{ineq00}? It is certainly true nowadays when the curvature is as small as $R_4\sim 12H_0^2\sim 10^{-60}M_P^2$. During the inflation, inequality \eqref{ineq00} also remains true. Indeed, for the chosen parameter set $R_4 =12 H^2\sim 10^{-11}M_P^2\sim 10^{-10}$ is much smaller than $R_3=\phi_0= 0.01$ during inflation.

Our analysis was performed on the basis of the Einstein frame. Meantime, there is no firm opinion which frame is realized in the Nature the Einstein frame or the Jordan one. In the latter case, we have to turn to equation \eqref{actTailor} after the evolution of the extra space is finished and the Ricci scalar $R_3=\phi_0$. The expression for the 4-dim Planck mass can be obtained by equating the value $M_P^2/2$ to the multiplier to $R_4$
$${M_P^2}=v_3e^{3\beta_0}f'(\phi_0)m_D^{2}.$$
Here expression \eqref{phi2} at the potential minimum $\phi=\phi_0$ is taken into account and the parameter $m_D$ is restored. For chosen parameter values $M_P\simeq 700m_D$ and previous estimations based on the Einsten frame remains the same.

We conclude that the compact space volume could remain small enough during the whole period of its evolution - from its nucleation at the (sub-)Planckian energies up to the modern epoch.

\section{Quantum fluctuations and stability of extra space}

In this section, we shortly discuss the role of quantum fluctuations on the stability of the extra space metrics discussed above. We start with the low energy scale $M_{low}$ where the present horizon is formed.
The quantum fluctuations of the scalar field $\phi$ have been intensively studied \cite{1987NuPhB.282..555K,1992JETPL..55..489S}. The common conclusion is that in spite of their smallness they are the reason of the large scale structure formation in our Universe. The amplitude of the fluctuations during the slow roll inflation is of the order of the $\delta\phi\sim H/2\pi\sim 10^{-7}M_P$.
They are also very small in the $m_D$ units: $\delta\phi\sim 10^{-6}$.

In this research, the Ricci scalar $R_3(\beta)$ plays the role of the scalar field $\phi$ - the inflaton - with the potential \eqref{VE}. Estimations made above indicate that the inflationary stage is performed near the bottom of the potential far from the potential extrema, see FIG. (\ref{pot}). Therefore we may not worry about the influence of the fluctuations on the final state of inflaton. If the initial value of the inflaton is to the right of the maximum, it will inevitably move to nonzero minimum. This indicates the stability of the metric evolution $\beta(t)=-\frac12 \ln [R_3(t)/6]=-\frac12 \ln[\phi(t)/6]\rightarrow -\frac12 \ln[\phi_{end}/6]$  with respect to the radial quantum fluctuations at the inflationary scale of energy.

The problem arises when we shift the scale up to the sub-Planckian energies where the quantum fluctuations $\delta\beta$ are of the order of the unity. Their role is twofold. If the region of stability is small, the quantum fluctuations could easily break the stationary behaviour of the extra space. The latter starts expanding or shrinking so that the probability of staying in a stable region is very small. On the other side, suppose that there exists the set of parameters $a,k,c$ leading to quasi stable classical solutions with slowly expanding sub-spaces $W_3$. Quantum fluctuations are able to turn its metric back to a stationary regime in some causally connected domains of the large sub-space $V_3$. Such domains could survive up to the beginning of the inflation.



\section{Conclusion}

The appearance of manifolds with different metrics as a result of quantum effects at high energies is a well known paradigm. After their creation, some manifolds evolve classically. The originated metrics serve as the initial conditions for their subsequent classical evolution.
The measure of any metric originated from space-time foam is assumed nonzero though uncertain due to the absence of the Theory of Quantum Gravity.

The analysis performed in this paper indicates that there are several regimes of sub-spaces evolution at sub-Plankian energies.
There are regimes characterized by the expansion of both sub-spaces at equal rates as well as at different rates depending on the Lagrangian parameters. We also have shown that some sub-spaces come back to the space-time foam.

The observable fact is that the only one 3-dim sub-space is large. It is those space where the modern physical processes are performed. Therefore, the regime characterized by only one growing sub-space is of most interest. We have found that such a regime is realized  at the highest energies for specific values of the Lagrangian parameters $a(m_D), k(m_D),c(m_D)$ and the specific initial metrics.

The parameter values of the Lagrangian $a(M_{infl}),k(M_{infl}),c(M_{infl})$ at the inflationary scale $M_{infl}$ were also discussed on the basis of chaotic inflation. They do not coincide with those at high
energies $m_D$ due to uncontrolled quantum corrections at the sub-Planckian energy scale.

Shortly, the general picture is as follows. Sub-spaces are nucleated with different initial metrics.  There is a class of multidimensional models with the higher derivatives for which some of the sub-spaces form pairs evolving classically in proper manner -  one of the sub-spaces expands while the other remains constant.

\section*{Acknowledgments}

The work is performed according to the Russian Government Program of Competitive Growth of Kazan Federal University and MEPhI Academic Excellence Project (contract \textnumero~02.a03.21.0005, 27.08.2013). The work of A.P. was  also supported by the Russian Foundation for Basic Research Grant No 19-02-00496. The work of S.G.R. was also supported by the Ministry of Education and Science of the Russian Federation, Project \textnumero~3.4970.2017/BY.


\end{document}